# GPGPU Based Parallelized Client-Server Framework for Providing High Performance Computation Support

[1]**Poorna Banerjee,** [2]**Amit Dave**
[1]Nirma Institute of Technology, Ahmedabad, India
[2]Space Applications Center, ISRO, Ahmedabad, India

## Abstract
Parallel data processing has become indispensable for processing applications involving huge data sets. This brings into focus the Graphics Processing Units (GPUs) which emphasize on many-core computing. With the advent of General Purpose GPUs (GPGPU), applications not directly associated with graphics operations can also harness the computation capabilities of GPUs. Hence, it would be beneficial if the computing capabilities of a given GPGPU could be task optimized and made available.
This paper describes a client-server framework in which users can choose a processing task and submit large data-sets for processing to a remote GPGPU and receive the results back, using well defined interfaces. The framework provides extensibility in terms of the number and type of tasks that the client can choose or submit for processing at the remote GPGPU server machine, with complete transparency to the underlying hardware and operating systems. Parallelization of user-submitted tasks on the GPGPU has been achieved using NVIDIA's Compute Unified Device Architecture (CUDA).

## Keywords
Client-Server Framework, High Performance Computation Support, GPGPU

## I. Introduction
Over the past few decades, microprocessor design has shown tremendous improvements, with the surfacing of two main design trajectories:
- The multi-core trajectory seeks to maintain the execution speed of sequential programs while moving into multiple cores. Most general computing desktops, workstations, etc. have multi-core microprocessors in their Central Processing Units (CPU), with the number of cores ranging from two upwards.
- The many-core trajectory focuses more on the execution throughput of parallel applications.

The many-core microprocessor design is exemplified by Graphics Processing Units (GPU) such as the NVIDIA's GeForce and Tesla series, AMD's Northern Island series, ATI Radeon R300 series, to name a few. Since 2003, many-core processors, especially GPUs, have led the race in floating point performance, when compared with their multi-core counterparts [1], as illustrated in fig. 1.
With the advent of general purpose programming on GPUs (GPGPU), even those applications which are not directly associated with graphics operations can still utilize the computing capabilities of graphic processing units. The Compute Unified Device Architecture (CUDA) introduced by NVIDIA helps in simplifying the interface to GPGPU programming. With today's scientific and even many commercial applications requiring very large data-sets for processing, parallel processing of data has literally become indispensable, since sequential processing of such huge data-sets is practically unviable. The processing of such huge data-sets often involves highly compute-intensive tasks and usually a group of such compute-intensive tasks have to be performed one after the other, on such data-sets. Common examples would include Look-up Table (LUT) generation and Image Correction, Graphic Visualization, matrix multiplication operations, to list a few.

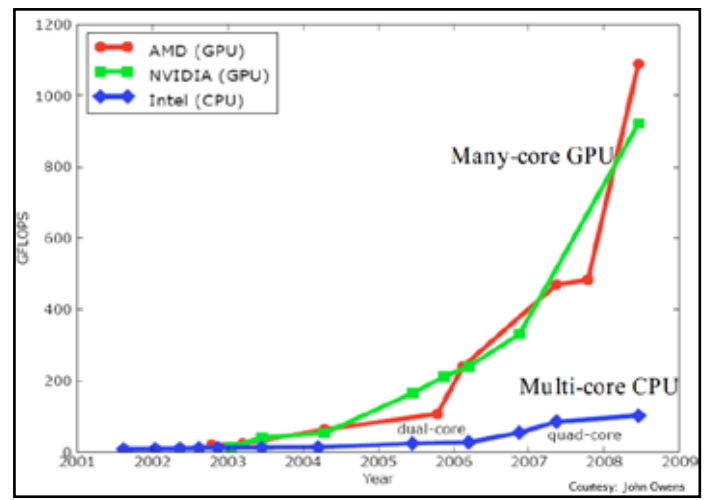

Fig. 1: Enlarging Performance Gap Between GPUs and CPUs [1]

Remarkable improvements are also surfacing on the network-performance front, with Gigabit networks, 10G and even 40G network backbones being adopted by organizations for fast and efficient transfer of data.
Keeping such factors in mind, the immediate benefit of a client-server framework, where users can choose a processing task and fire data-processing commands to a remote GPGPU server, becomes evident. The round-trip time incurred in transfer of data to and from the server machine, gets more than compensated by the tremendous speed-ups obtained in the performance through task parallelization on the GPGPU. In the proposed framework, the parallelization of user submitted processing tasks on the GPGPU has been implemented using NVIDIA's CUDA programming model.

## II. Client-Server Framework Design
The objective of the client-server framework presented in this paper is to support GPGPU based high performance computation with transparency to underlying hardware and operating systems, which would allow clients or users to fire data processing tasks to a remote GPGPU. The proposed system would provide dynamic extensibility in terms of the number and type of tasks the client can submit for processing at the remote server associated with the GPGPU, making it easy for future addition of more processing tasks to the already existing task-set. Development of such a framework involved establishment of client and server interfaces and management of distributed data structures. The user application interacts with client-side library modules which in turn interact with server-side modules associated with the GPGPU for data processing. The processed results are then sent back to the client. The basic scenario is demonstrated in fig. 2.

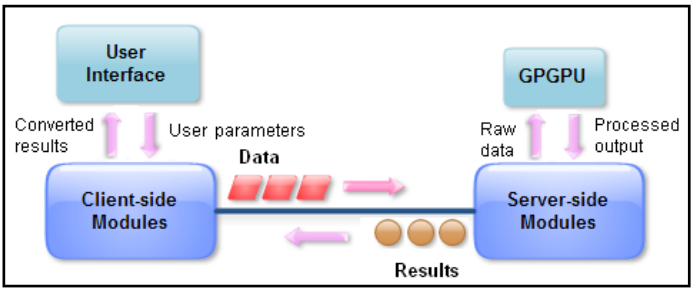

Fig. 2: Schematic Diagram for the Basic Client-Server Framework Scenario

Certain design issues came into perspective while developing this framework:
- Type of application interface that is to be provided to the client - it can be GUI or Command Line (CLI) based. Both GUI and CLI based support has been provided during the actual implementation of the framework.
- Various parameters that are needed to be passed to the client-side processing modules. Some such parameters could be - IP-address of the remote GPGPU server, input data file to be processed, output file in which the results are to be returned back, task flags and other options.

Representation of the data for transmission over the network. Certain aspects that need to be decided are - fields and format of the data sent and received, maximum allowable data size, required meta-data to be added such as data-type indicators, matrix-dimensions, etc.

One such template header structure, for sending client parameters to the server, is shown in fig. 3. Field sizes indicated in the header are in bytes. Here, the first field of 29 bytes in the header indicates the task flag that is generated when a client chooses a particular processing task. This task flag helps the GPGPU sever decide what kind of processing is to be carried out on the input data.

The next field is a one-byte special symbol which indicates to the server whether it should expect any input data from the client. A '+' indicates input data follows the header. A '\0' indicates absence of input data from the client. The next 200 bytes in the header have been allocated for specifying a comma separated list of parameters. The set of parameters present in the list depends on which particular task the client has chosen. Some common parameters include data-matrix dimensions, data type indicator, polynomial orders, etc. The next 30 bytes in the header specify the name of the output file in which the client wants to receive the results. All the fields in the header get populated whenever the client fires a task for processing, either via the GUI or CLI, and accordingly a suitable header structure is generated by the client-side processing modules and sent to the server.
- Technology that should be used to establish the connection and transmit data between the client and server. Some possible choices include Socket Programming, RPC, Java RMI, and CORBA. While the more sophisticated option such as CORBA and RMI work best with Java and introduce a layer of middleware functionalities, options such as RPC and socket-programming are more primitive, C/C++ based and can provide better performance. During actual implementation of the framework, C/C++ sockets have been used.
- Upon reception of processed results, the client-side modules would have to re-assemble the data to present it in an understandable form to the user application. This requires data conversion methods.
- Choice of transport layer protocol. Common choices would be TCP, UDP and XTP. While TCP is connection oriented and reliable, UDP is connectionless, unreliable but faster. Since most of the data transfer involved in the given client-server scenario involves data files which must be sent and received without any errors, TCP was chosen as the transport layer protocol.

Thus, designing the framework involved:
- Development of customized data exchange mechanisms between the client and server.
- Designing client-side and server-side modules for dynamic loading and unloading of user-submitted tasks and data.
- Development of efficient, parallelized programs for optimized processing at the GPGPU.

After establishing the system-framework, the following three functions to establish the capability of the framework and an essential utility for the user have been developed:
1. Bilinear Interpolation based Demosaicing of an Image.
2. Gradient Interpolation based Demosaicing of an Image.
3. Least-Squares Curve Fit for any Polynomial Order on Given Data.
4. Remote GPGPU Information Generation

A user-friendly GUI has been developed for the client so that various input parameters can be specified, as shown in fig. 4. The GUI has been designed so as to facilitate both command-prompt based and UI based execution of tasks submitted by the client. Additional features have been provided in the GUI for fault detection and error-log archiving.

| 0 | 28 | 29 | 30 | 229 | 230 | 259 | 260 ... ~ |
|---|---|---|---|---|---|---|---|
| Function name or task-flag (To indicate the processing task to be carried out) | | + (Special character to indicate whether end of header is followed by data) | Comma separated parameter string (Parameters selected by the client via the GUI) | | Output file name (Name of the file in which the GPGPU-server return results) | | Data (Input data file sent by the client to the GPGPU-server) |

Fig. 3: Header Template for Sending User Parameters to the Server (Field Sizes in Bytes)

In order to produce the demosaiced image, the process of interpolation has to be performed. There are many methods for performing such interpolation, two of them being Bilinear Interpolation and Gradient-Based Interpolation, which have been implemented on the client-server framework and as a case-study, details for Bilinear interpolation based image demosaicing has been discussed in the following subsection.

### 1. Bilinear Interpolation Based Demosaicing

In this method of interpolation, we calculate the average on each pixel depending on its position in the Bayer Pattern for determining the intensity of that particular pixel. For each pixel, we consider its 8 direct neighbors and then we determine the 2 missing color components of that pixel by averaging the corresponding colors of the neighboring pixels [3]. Depending on pixel position, we have 4 different cases of averaging: the pixel is red, the pixel is blue, the pixel is green and is in a blue row or the pixel is green and is in a red row. This method of interpolation allows us to determine the red, green and blue color component values for each pixel and compared to simple pixel-doubling produces better results, since the average of the neighbor pixels is calculated instead of just replicating the values. The missing color-component values will also be closer to the original ground-truth values.

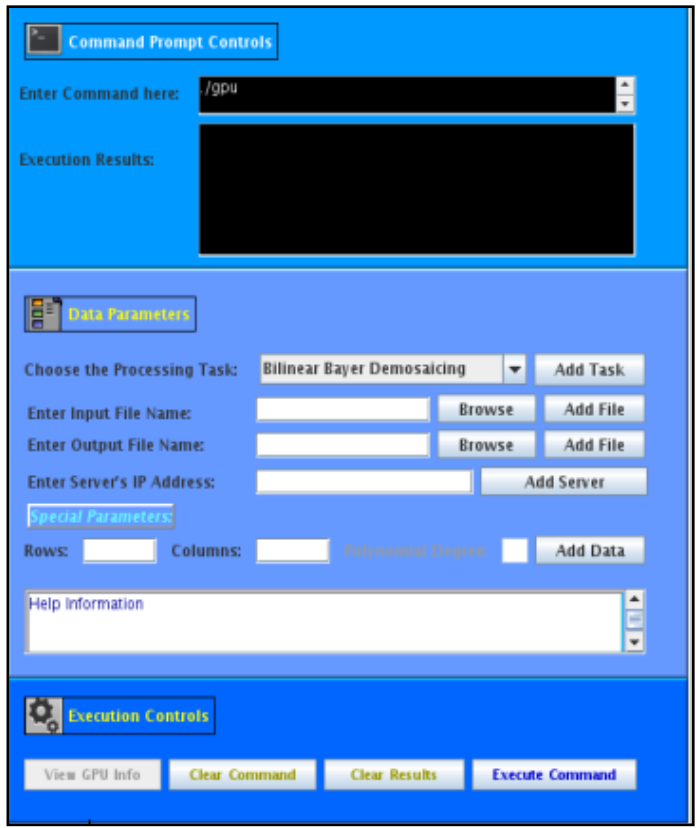

Fig. 4: GUI Interface for GPGPU Based Computation Support

### III. GPGPU-Based Parallelization of Tasks Submitted by the Client

This section has been organized as a group of case-studies, with each case-study describing a particular client submitted task, its parallelization on the GPGPU, along with the corresponding execution performance analysis. The first case-study describes image demosaicing, which involves ordinary arithmetic operations such as addition and averaging. The second case-study describes Least Square Errors curve fitting, which is mathematically complex and involves many compute intensive matrix operations.

### A. Case-Study 1: Image Demosaicing

Single-sensor imaging devices such as digital cameras, are based on a Charge Coupled Device (CCD) array or CMOS imaging sensor, with each pixel on the sensor capturing only one sample of the color spectrum [3-4]. Digital color cameras generally use a Bayer mask over the CCD. A Bayer filter mosaic is a Color Filter Array (CFA) for arranging Red Green Blue (RGB) color filters on a square grid of photo-sensors, as shown in fig. 5. Each square of four pixels has one filtered red, one blue, and two green pixels. This is because the human eye is more sensitive to green than either red or blue and the Bayer CFA is based on the knowledge of human visual perception.

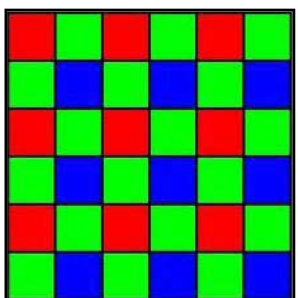

Fig. 5: The Bayer Pattern [6]

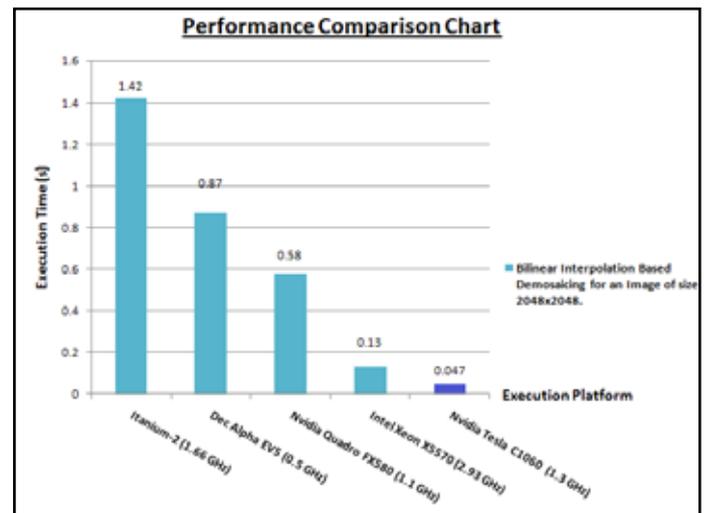

Fig. 6: Performance Chart for Bilinear Interpolation Based Demosaicing

### 2. Execution Performance Analysis

The task of Bilinear interpolation based demosaicing was performed on images of size 2048x2048, with each pixel's intensity value being represented by 16 bits. Performing demosaicing on an image using Bilinear interpolation requires similar actions to be performed on each pixel of the image, and these operations can be carried out on each pixel simultaneously. This provides a very good scope for parallelizing the task of demosaicing an image using Bilinear interpolation.

In the client-server framework, any client may submit raw image data to the remote GPGPU server for performing Bilinear interpolation based image demosaicing via the GUI; the client needs to specify the IP address of the server, name of the input image-data file, the type of processing task to be carried out (in this case, its Bilinear Bayer Demosiacing) and the name of the output-file in which the results are to be returned. These controls can be specified either via the command prompt controls or through the GUI based data parameters. Upon successful execution and thereafter completion of the submitted task, the demosaiced image is sent back to the client in the specified output file name. A parallelized, CUDA/C based implementation has been developed which has been tested on the NVIDIA Tesla C1060 GPGPU and NVIDIA Quadro FX580 graphics-card. The sequential version of the same program has been implemented and tested on various platforms. A comparative performance chart, along with the corresponding speed-ups obtained, are shown in fig. 6 and Table 1 respectively. As a sample output, the mosaiced input image and the demosaiced output images are shown in fig. 7. Fig. 7(a) shows the original mosaiced image, fig. 7(b) shows the demosaiced output image. Fig. 7(c) shows a zoomed out portion of the original image to highlight the mosaiced checkerboard pattern. Fig. 7(d) shows the same demosaiced portion from the output image.

Comparison Benchmark:
NVIDIA Tesla C1060
Compute Capability: 1.3
Clock Rate: 1.3 GHz
No. of Processor Cores: 240
Total Global Memory: 4GB
Speed-up obtained over test platform =

$$\frac{Execution\ time\ on\ Test\ Platform}{Execution\ time\ on\ Comparison\ Benchmark}$$

Table 1: GPGPU Speed-Ups Obtained from Parallelizing Bilinear Interpolation Based Image Demosaicing

| Name of Test Platform | Test Platform Specifications | Speed-up obtained over this Platform |
|---|---|---|
| Itanium-2 | Model: 1  Architecture: IA-64  Clock Rate: 1.66 GHz | 30x |
| Dec Alpha EV5 | Model: 21164  Architecture: DEC 64-bit RISC  Clock Rate: 0.5 GHz | 18x |
| NVIDIA Quadro FX580 | Compute Capability: 1.1  Clock Rate: 1.1 GHz  No. of Processor Cores: 32  Total Global Memory: 0.5 GB | 12x |
| Intel Xeon | Model: X5570  Architecture: Nehalem-EP, x86-64  Clock Rate: 2.93 GHz | 3x |

### B. Case-Study 2: Least Square Errors Curve Fit for any Polynomial Order on Given Data

#### 1. Estimation of Best-Fit Curve Using Least Square Errors Method

The procedure of Least Squares curve-fitting is exquisitely used in many applications for fitting a polynomial curve of a given degree to approximate a set of data. If the input data-set is represented by n data-point pairs of the type $(x_i, y_i)$ where $1 \leq i \leq n$, $n \geq 2$, then the best fitting curve $f(x)$ has the least square error [5], i.e.,
$\Pi = \Sigma[y_i - f(x_i)]^2$ = minimum, where i = 1 to n.

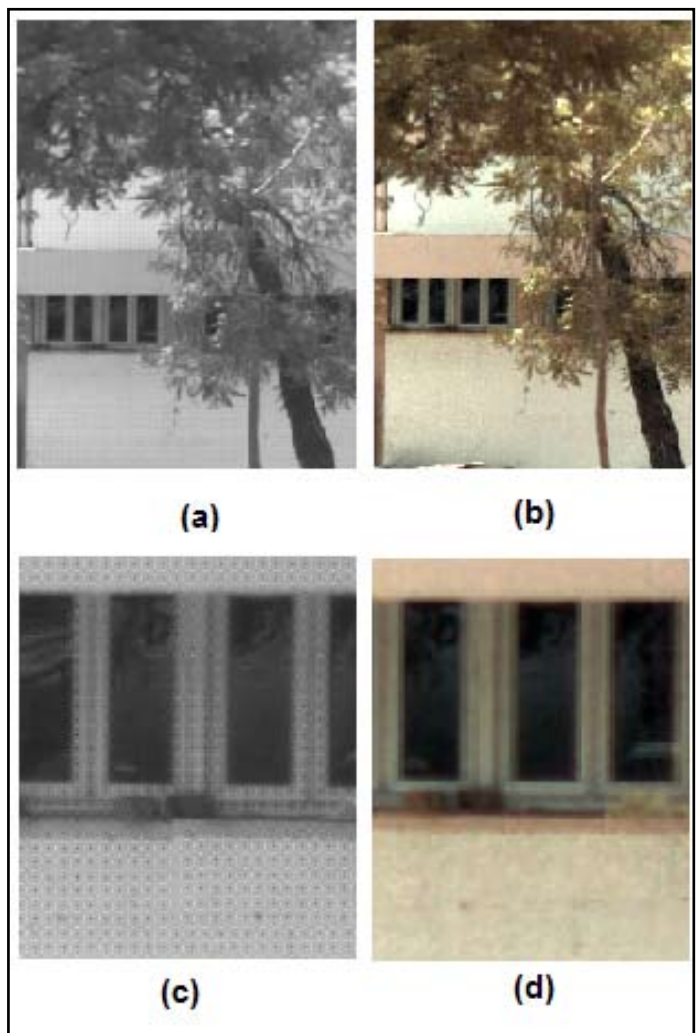

Fig. 7: Results of Bilinear Interpolation Based Image Demosaicing

For a $m^{th}$ degree polynomial fit:
$f(x) = a_0 + a_1 x + a_2 x^2 + ... + a_m x^m$, where $a_1, a_2, ..., a_m$ are coefficients to be determined.
To obtain the least square error, the unknown coefficients must yield zero first derivatives, which lead to the following equations:
$\delta \Pi / \delta a_j = 2\Sigma[y_i - (a_0 + a_1 x + a_2 x^2 + ... + a_m x^m)] = 0$,
where i = 1 to n, j = 0 to m
Expanding the above set of equations, we get:
$\Sigma x_i^j y_i = a_0 \Sigma x_i^j + a_1 \Sigma x_i^j+1 + ... + a_m \Sigma x_i^{j+m}$,
where i = 1 to n, j = 0 to m.
The unknown coefficients $a_0, a_1, ..., a_m$ can hence be obtained by solving the above set of linear equations. The set of equations described above indicate that in order to determine the unknown coefficients $a_0, a_1, ..., a_m$, we have to solve a system of linear equations of the form AX = B, where the matrices A, X, B are given by:
Matrix A:

$\begin{array}{cccc} 1 & \Sigma x_i & \Sigma x_i^2 & ... & \Sigma x_i^m \\ \Sigma x_i & \Sigma x_i^2 & \Sigma x_i^3 & ... & \Sigma x_i^{m+1} \\ ... & & & & \\ \Sigma x_i^m & \Sigma x_i^{m+1} & \Sigma x_i^{m+2} & ... & \Sigma x_i^{2m} \end{array}$

where i = 1 to n.
Matrix X:
[ $a_0$  $a_1$  $a_2$  ...  $a_m$ ]

Matrix B:
$[\ \Sigma y_i\ \ \Sigma x_i y_i\ \ \Sigma x_i^2 y_i\ ...\ \Sigma x_i^m y_i\ ]$
where i = 1 to n.
The matrix X can now be solved by evaluating the inverse of matrix A, i.e. $X = A^{-1}B$.

### 2. Execution Performance Analysis

The procedure of curve fitting was applied on an experimentally obtained data-set with 6 scan lines and 6000 pixels per scan-line. The sequential versions as well as the parallel versions of the curve-fitting programs were implemented on many platforms, and a performance comparison chart is shown in fig. 8. From the performance chart, it is evident that tremendous speed-ups have been achieved. Speed-up's obtained, for varying polynomial orders, are summarized in Table 2.

Table 2: GPGPU Speed-Ups Obtained from Parallelizing Least Square Errors Curve Fitting

| Name of Test Platform | Test Platform Specifications | Speed-up obtained over this Platform |
| --- | --- | --- |
| Itanium-2 | Model: 1<br>Architecture: IA-64<br>Clock Rate: 1.66 GHz | Order 1: 47x<br>Order 2: 66x<br>Order 3: 73x |
| Dec Alpha EV5 | Model: 21164<br>Architecture:<br>DEC 64-bit RISC<br>Clock Rate: 0.5 GHz | Order 1: 11x<br>Order 2: 37x<br>Order 3: 45x |
| Intel Xeon | Model: X5570<br>Architecture:<br>Nehalem-EP, x86-64<br>Clock Rate: 2.93 GHz | Order 3: 18x |

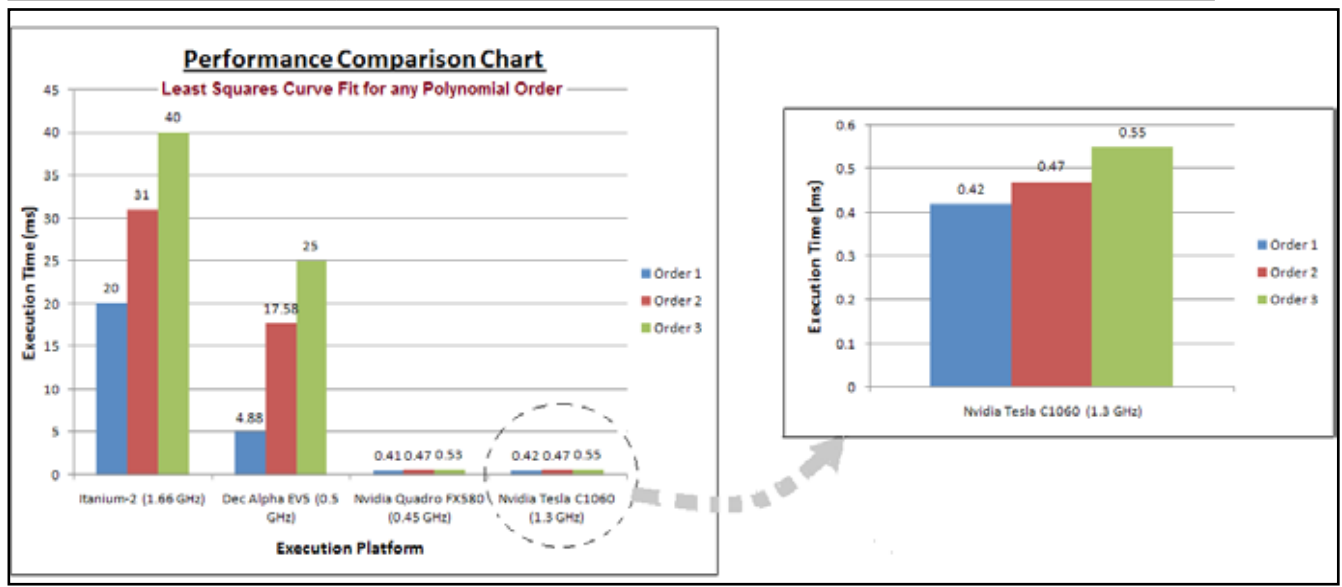

Fig. 8: Performance Chart for Least Square Errors Curve Fitting

### IV. Framework Utilities and Up- Gradation

The module for remote GPGPU information generation serves as a utility to the client, which when executed gives the client all the information about the various GPU resources available on the Remote GPGPU server. A complete listing with all the specifications of each GPU resource is generated in the form of an XML file which is then sent back to the client. On the GUI, the client can view this data in the form of a tree-structure with associated attributes. The client only needs to specify the GPGPU server and the name of the output file. The resultant XML file is sent back to the client and this data can be viewed in the form of an XML-tree in the GUI. Various GPU specifications include:
- Name , Compute Capability
- Warp Size
- Total Constant Memory Size
- Total Global Memory Size

- Shared Memory Per Block
- Clock Rate
- Multi Processor Count
- Registers Per Block
- Maximum Threads Per Block
- Maximum Grid Size
- Maximum Threads Dimensions

When adding further tasks to the framework is considered, the framework has been designed keeping in mind that future developers can contribute their GPGPU library codes to the already existing task-set. The only steps that code developers have to follow is to create their library codes according to a certain generic template designed specially for the framework, following which the new libraries can be seamlessly integrated with the existing task-set through creation of shared, dynamically loaded libraries. All that is needed is a one-step compilation to generate the shared-library object.

## V. Conclusions and Future Scope of Work

With evolving requirements of today's scientific applications, the processing paradigm has shifted from multi-core to many-core processing which brings into focus the GPGPUs which are ideal for supporting compute-intensive, parallel operations. This paper describes a platform-independent client-server framework to support GPGPU based high performance computation with transparency to underlying hardware and operating systems. Task parallelization on the GPGPU has shown tremendous speed-ups in execution performance compared to the sequential versions of the same.

As future scope for work, the framework can be extended to perform more parallel processing tasks such as carrying out Modulation Transfer Function (MTF) for real-time data. MTF is a parameter used for measuring the sharpness of an imaging system. The data-set would be in the range of a few gigabytes. Processing such a huge amount of data has associated issues such as:

- I/O time taken up in transferring the data to and from the client/server.
- I/O time taken up in transferring the data to and from the host CPU to the GPGPU.

For example, on a gigabit network, transmitting a typical MTF data file with size 2.5GB would itself take 20 seconds! In order to reduce the transmission latency incurred, one option is to apply lossless data-compression techniques to the data. One solution for minimizing the latency incurred in transferring data between the CPU and the GPGPU is to use memory-mapped I/O, where the GPU and the host CPU would share the same memory address space. This would eliminate the need for copying data frequently to and from the GPU to the CPU. NVIDIA GPUs with compute capability 2.x or higher possess such a feature.

Also, the utility for generating information about remote GPGPU resources can be extended to enable the client to choose the GPGPU resource on which he or she wants to execute the chosen task. This would involve associating resource allocation algorithms with the framework to allocate the available GPGPU resources in the best possible way.

## VI. Acknowledgement

We sincerely acknowledge the constant motivation and guidance from Mr. Ashish Mishra, Head (Payload Checkout Software and Vision Systems Division, PCEG/SEDA) and Mr. D. R Goswami, Group Director (PCEG/SEDA). We would also like to extend our heartfelt gratitude to Dr. Madhuri Bhavsar, Section Head (IT), Nirma Institute of Technology, for her continual support and guidance.

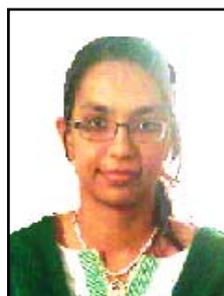

Poorna Banerjee has received her B.Tech Degree in Computer Science and Engineering from Nirma Institute of Technology, Ahmedabad, India, in 2010 and is pursuing M.Tech in Computer Science and Engineering in the same institute and is currently engaged in final year dissertation at Space Applications Center, ISRO, Ahmedabad, India. Her research interests include image processing, high performance computing, parallel processing and wireless sensor networks.

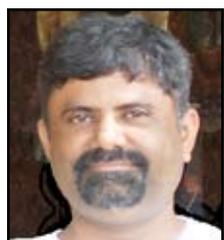

Amit Dave received his B.E. degree in Computer Science from Faculty of Technology and Engineering, M. S. University, Vadodara, Gujarat, India in 1992. He is associated with Indian Space Research Organization as Scientist/Engineer. He has developed software in different areas like device drivers, instrumentation, automation, database and scientific applications for payload checkout segment of various Indian Remote Sensing cameras. His research interest includes grid/cluster computing, high performance computing and high speed data acquisition. He served as Project Manager (Payload Checkout Software) for Chandrayaan-1 mission. He is presently engaged in the development of next generation payload checkout software for future missions.